\documentclass{article}
\usepackage{spconf,amsmath,graphicx}
\usepackage{amsfonts}
\usepackage{algorithmic, algorithm}
\usepackage{dsfont, braket, amsmath, amssymb, textcomp, mathtools}
\usepackage{placeins}
\usepackage[table]{xcolor}
\usepackage{booktabs}       
\usepackage{multirow}
\usepackage{makecell}
\usepackage{hyperref}
\usepackage{colortbl}
\usepackage{subcaption}
\usepackage[percent]{overpic}
\usepackage{setspace}

\def\obs{{\mathbf y}}
\def\img{{\mathbf x}}
\def\noi{{\mathbf \epsilon}}

\newtheorem{finding}{Finding}

\newcommand{\bm}[1]{{\mathbf#1}}
\renewcommand{\paragraph}[1]{{\noindent\bf#1}}

\newcommand{\Section}[1]{\vspace{-10pt}\section{#1}\vspace{-10pt}}
\newcommand{\Subsection}[1]{\vspace{-14pt}\subsection{#1}\vspace{-7pt}}

\title{D-VDAMP: Denoising-based Approximate Message Passing\\ for Compressive MRI}
\name{Christopher A. Metzler, Gordon Wetzstein \thanks{C.M.~was supported by an appointment to the Intelligence Community Postdoctoral Research Fellowship Program at Stanford University administered by Oak Ridge Institute for Science and Education (ORISE) through an interagency agreement between the U.S. Department of Energy and the Office of the Director of National Intelligence (ODN). G.W.~was supported by an NSF CAREER Award (IIS 1553333), a Sloan Fellowship, and a PECASE by the ARL.}}
\address{Department of Electrical Engineering at Stanford University\\
	\texttt{\{cmetzler,gordon.wetzstein\}@stanford.edu}}

\begin{document}

\setlength{\abovedisplayskip}{3pt}
\setlength{\belowdisplayskip}{3pt}
\setlength{\abovedisplayshortskip}{0pt}
\setlength{\belowdisplayshortskip}{0pt}

%
\maketitle
\begin{abstract}

	Plug and play (P\&P) algorithms iteratively apply highly optimized image denoisers to impose priors and solve computational image reconstruction problems, to great effect. However, in general the ``effective noise'', that is the difference between the true signal and the intermediate solution, within the iterations of P\&P algorithms is neither Gaussian nor white. This fact makes existing denoising algorithms suboptimal.
	
	In this work, we propose a CNN architecture for removing colored Gaussian noise and combine it with the recently proposed VDAMP algorithm, whose effective noise follows a predictable colored Gaussian distribution. We apply the resulting denoising-based VDAMP (D-VDAMP) algorithm to variable density sampled compressive MRI where it substantially outperforms existing techniques.  

\end{abstract}
\Section{Introduction}
\label{sec:intro}

The goal of computational imaging (CI) is to recover a vectorized image $\img_o\in\mathbb{C}^n$ from a measurement $\obs$. 
For the special case of compressive magnetic resonance imaging (MRI), which is one of the more mature and important applications of CI, the forward model can be approximated by
\begin{align}
\obs=\bm{M}_\Omega( \bm{F}\img_o+\noi),
\end{align}
where $\bm{F}\in\mathbb{C}^{n\times n}$ is a 2-D discrete Fourier transform matrix, $\bm{M}_\Omega\in\mathbb{R}^{n\times n}$ is a diagonal matrix, parameterized by $\Omega\in\mathbb{R}^n$, that models subsampling the Fourier coefficients, and $\noi$ represents complex circular Gaussian distributed noise. The elements $\bm{M}_{i,i}$ are one if $\Omega_i=1$ and zero otherwise. There are $m\ll n$ non-zero elements in $\Omega$. 

The quality of any CI imaging system, including MRI, depends on its recovery algorithm. In the last two decades, dozens of CI recovery algorithms have been developed.  Among them, plug and play (P\&P) approaches~\cite{BM3DasReg,PnPADMM,metzler2016denoising,RED}, which iteratively apply a denoising algorithm to impose a prior on the reconstruction, have generally been the most successful. P\&P approaches using convolutional neural networks (CNNs) have been particular effective~\cite{LDAMP}. For a comprehensive review of plug and play methods applied to compressive MRI see~\cite{ahmad2019plug}.

Denoising-based approximate message passing (D-AMP) \cite{metzler2016denoising,donoho2009message,schniter2016denoising} is an especially exciting P\&P algorithm. It converges much faster than competing algorithms and comes with a {\em scalar} state evolution framework that allows one to predict and reason about its performance. These strengths are a result of a property unique to AMP algorithms: The difference between the ground truth signal and the intermediate reconstructions in AMP, a term known as the ``effective noise'', can be modeled as additive {\em white} Gaussian noise with a known variance~\cite{donoho2009message}.

However, like the AMP algorithm on which it is based, for this property to hold D-AMP requires that the elements of the forward operator/measurement matrix follow an i.i.d.~sub-Gaussian distribution. When the elements of the matrix are not i.i.d.~sub-Gaussian distributed, the effective noise is not white, the state evolution framework does not predict the reconstruction accuracy, and the performance of the algorithm suffers~\cite{AMPwArbitraryMatrices,caltagirone2014convergence}. 

Much work has gone into extending AMP to work with other measurement matrices. Through the use of damping~\cite{AMPwArbitraryMatrices}, sequential updates~\cite{caltagirone2014convergence,SwAMP}, variable splitting~\cite{VAMP}, and divergence-free denoisers~\cite{OAMP}, AMP can be extended to work with a far broader class of matrices. However, none of the aforementioned extensions fully support Fourier measurements, which are arguably the most important CS measurement operator. As a result, to date AMP's performance on Fourier-based CS problems like MRI has been underwhelming~\cite{eksioglu2018denoising}.

Recently, Millard et al.~\cite{millard2020approximate} proposed a novel extension to AMP, coined variable density AMP (VDAMP), in which the reconstruction is produced and predicted on a per wavelet subband basis (related ideas were proposed in~\cite{philsposter}). 
Under this model, the effective noise at each iteration is additive {\em colored} Gaussian noise with a covariance matrix which is diagonal when the noise is represented in the wavelet domain. Moreover, VDAMP comes with a (weak) {\em vector} state evolution framework that can predict the variance of the effective noise in each subband, even when dealing with Fourier measurements. 
Still, as will be demonstrated in Section~\ref{sec:Sims}, the performance of VDAMP trails behind that of P\&P methods which can leverage far more sophisticated priors.

In this work, we propose, analyze, and test a P\&P denoising-based VDAMP (D-VDAMP) algorithm which extends and substantially improves the performance of VDAMP by incorporating a novel CNN-based colored noise removal algorithm in place of soft wavelet thresholding.

\Section{D-VDAMP}

The proposed D-VDAMP algorithm is presented in Algorithm~\ref{alg:DVDAMP} and is described below. 
Like the VDAMP algorithm on which it is based, D-VDAMP forms a series of noisy ($\bm{r}_k$) and denoised ($\hat{\bm{w}}_k$) estimates of the true signal's wavelet coefficients, $\bm{w}_o=\bm{\Psi}\img_o$, where $\bm{\Psi}$ denotes a forward wavelet transform~\cite{millard2020approximate}. We use a four level 2-D Haar transform throughout the paper.

\begin{algorithm}[t]
	\caption{D-VDAMP\label{alg:DVDAMP}}
	\textbf{Input:} Sampling set $\Omega$, wavelet transform $\bm{\Psi}$, probability matrix $\bm{P}$, measurements $\bm{y}$, regularization parameter $\bm{\gamma}$, image domain denoiser $\bm{D}(\bm{\img}; \gamma\boldsymbol{\tau})$,  number of iterations $K_{it}$.
	\begin{algorithmic}[1]
		\STATE Set $\widetilde{\bm{r}}_{0} = \bm{0}$, set $\boldsymbol{\tau}_{-1}=\boldsymbol{\infty}$, and compute $\bm{S} = |\bm{F}\bm{\Psi}^H|^2$\label{algline:init}
		\FOR {$k =0,1,\ldots,K_{it}-1$} 
		\STATE $\bm{z}_k = \obs - \bm{M}_\Omega \bm{F} \bm{\Psi}^H\widetilde{\bm{r}}_{k}$ \label{algline:resid}
		\STATE $\bm{r}_{k} = \widetilde{\bm{r}}_{k} + \bm{\Psi} \bm{F}^H \bm{P}^{-1}\bm{z}_k$ \label{algline:pseudodata}
		\STATE $\boldsymbol{\tau}_{k} = \bm{S}^H\bm{M}_\Omega \bm{P}^{-1} [(\bm{P}^{-1} - \mathds{1}_N)|\bm{z}_k|^2 + \sigma_\epsilon^2 \bm{1}_N] $ \label{algline:WaveletVariance}
		\IF{$\|\boldsymbol{\tau}_k\|_1 > \| \boldsymbol{\tau}_{k-1} \|_1$}\label{algline:break_condition_1}
			\STATE \textbf{break}\label{algline:break_condition_2}
		\ENDIF\label{algline:break_condition_3}
		\STATE $\hat{\bm{w}}_{k} = \bm{\Psi}\bm{D}(\bm{\Psi}^H\bm{r}_{k}; \bm{\gamma}\boldsymbol{\tau}_{k})$ \label{algline:denoise}
		\STATE $\boldsymbol{\alpha}_{k} = \text{M.C. estimate of }\braket{\bm{\partial} \bm{\Psi}\bm{D}(\bm{\Psi}^H\bm{r}_{k}; \bm{\gamma} \boldsymbol{\tau}_{k})}_\mathrm{sband}$ \label{algline:divergence}
		\STATE $\widetilde{\bm{r}}_{k+1} = (\hat{\bm{w}}_{k} - \boldsymbol{\alpha}_k \odot \bm{r}_{k})\oslash (\bm{1}_N-\boldsymbol{\alpha}_k)$ \label{algline:tildeR}
		\ENDFOR
		\RETURN $\hat{\img} = \bm{\Psi}^H\hat{\bm{w}}_k$ 
		\label{algline:final_result}
	\end{algorithmic}
\end{algorithm}

On line~\ref{algline:resid} the algorithm computes the residual error between the observations $\bm{y}$ and the wavelet coefficient estimate $\widetilde{\bm{r}}_k$. 
On line~\ref{algline:pseudodata}, D-VDAMP uses this estimate of the residual to take a {\em density compensated} gradient descent step. That is, rather than taking a gradient step to reduce ${\|\bm{y}-\bm{M}_\Omega( \bm{F}\img)}\|^2$, the algorithm instead takes a step to reduce ${\mathbb{E}_{\Omega}\|\bm{y}-\bm{M}_\Omega( \bm{F}\img)}\|^2$. This step is central to whitening the effective noise within the algorithm but, because it involves scaling the residual by $\bm{P}^{-1}$, this modification makes the algorithm more sensitive to measurement noise~\cite{millard2020approximate}.

On line~\ref{algline:WaveletVariance} the algorithm computes an estimate of the variance in the wavelet domain.
On lines~\ref{algline:break_condition_1}--\ref{algline:break_condition_3} the algorithm compares the estimated power of the effective noise associated with the current iteration with the estimated power from the previous iteration. If the current iteration's estimate is larger, the solution is getting worse and the algorithm terminates.
On line~\ref{algline:denoise} the algorithm applies an image domain denoiser. The denoiser is parameterized by the estimated diagonal, $\boldsymbol{\tau}_{k}$, of the effective noise's covariance matrix and a tunable parameter, $\bm{\gamma}$, which determines the amount of regularization.
On line~\ref{algline:divergence} the algorithm computes a Monte Carlo estimate of the average partial derivative associated with denoising the elements of each wavelet subband. See Section~\ref{ssec:Divergence} for more information. 

Line~\ref{algline:tildeR} is the colored Onsager correction step. 
The intuition behind this step is that each wavelet band of the reconstruction is updated at a different rate. The lower the average partial derivative with respect to a subband (which is a proxy for the variance associated with the reconstruction in that sideband), the faster that band is updated. See~\cite{donoho2009message,millard2020approximate} for more information about the Onsager correction. 

Finally, on line~\ref{algline:final_result} the algorithm maps $\hat{\bm{w}}_{k}$ back to the image domain to form the final estimate of the image. Note, that unlike VDAMP, our algorithm does not have a final gradient descent step on line~\ref{algline:final_result}. The exclusion of this term significantly improved the algorithm's performance at low SNRs, while trading off some performance at higher SNRs. 

\Subsection{Calculating Average Partial Derivatives}\label{ssec:Divergence}
A key step in the VDAMP algorithm is computing the average partial derivative across each wavelet subband, $\braket{\bm{\partial} \bm{\Psi}\bm{D}(\bm{\Psi}^H\bm{r};\bm{\gamma} \boldsymbol{\tau})}_\mathrm{sband}$. In this work, we do so by extending the Monte-Carlo procedure developed in~\cite{MCSURE}. 

Let $g(\bm{r})=\bm{\Psi}\bm{D}(\bm{\Psi}^H\bm{r}; \bm{\gamma}\boldsymbol{\tau})$. For each subband $s$, we estimate the average partial derivative of the real part of $g(\bm{r})$ with
\begin{align}\label{eqn:MC_Div}
\braket{\bm{\partial} \mathcal{R}(g(\bm{r}))}_\mathrm{sband} &\approx \frac{1}{J_s} \mathcal{R}(\bm{b}_s^t) \left( \frac{\mathcal{R}(g(\bm{r}+\eta \mathcal{R}(\bm{b}_s))-g(\bm{r}))}{\eta} \right)\nonumber
\end{align}
where $\mathcal{R}(\cdot)$ denotes taking the real part of the argument, $\eta$ is a small constant (we use $\frac{|\max(\bm{r})|}{1000}$), $J_s$ is the number of elements in subband $s$, and $\bm{b}_s$ is a vector vector whose elements within the band $s$ follows an i.i.d. circular complex Gaussian distribution with variance $2$ (real and imaginary part each have variance $1$) and whose elements outside this band are $0$. 
An analogues expression is used to compute the average partial derivative of the imaginary component. 
Following~\cite{millard2020approximate}, we compute the average partial derivative of the complex function by averaging the partials from the real and imaginary parts.

\Subsection{Vector ``State Evolution''}

\begin{figure}[t]
	\centering
	\begin{subfigure}[b]{0.45\textwidth}
		\centering
		\begin{overpic}[width=.3\textwidth]{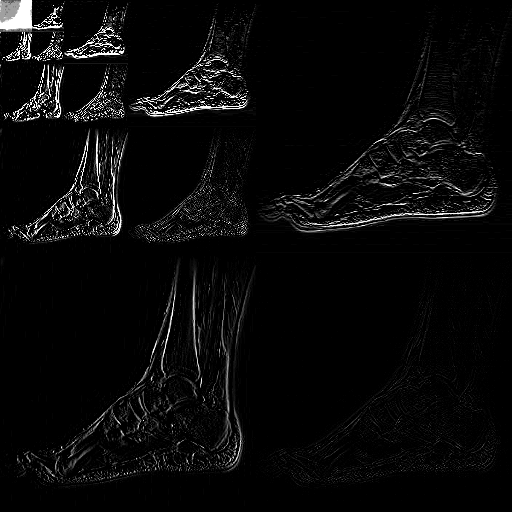}
			\put (45,-12) {{\centering $\bm{w}_o$}}
		\end{overpic}
		\begin{overpic}[width=.3\textwidth]{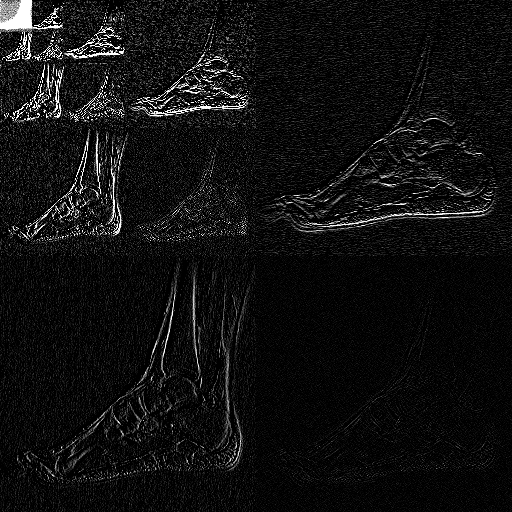}
			\put (45,-12) {{\centering $\bm{r}_2$}}
		\end{overpic}
		\begin{overpic}[width=.3\textwidth]{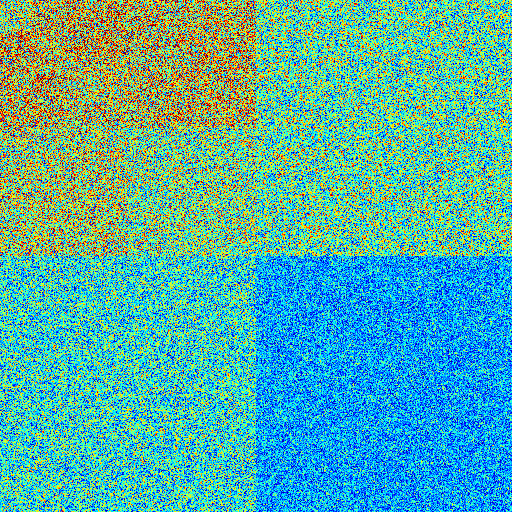}
			\put (25,-12) {{\centering $|\bm{r}_2-\bm{w}_o|$}}
		\end{overpic}
		\vspace{15 pt}
		\label{fig:ColoredNoise}
	\end{subfigure}
	\hfill
	\begin{subfigure}[b]{0.45\textwidth}
		\centering
		\includegraphics[width=\textwidth]{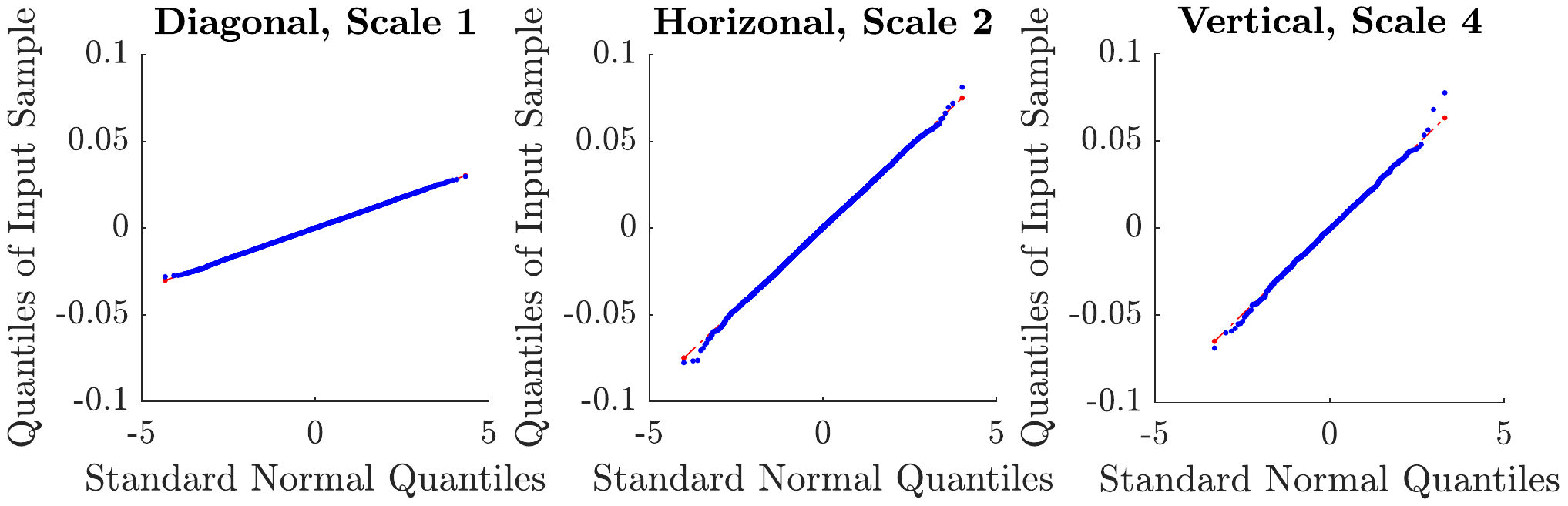}
		\vspace{-7 pt}
		\label{fig:QQplots}
	\end{subfigure}
	\hfill
	\begin{subfigure}[b]{0.5\textwidth}
		\centering
		\includegraphics[width=\textwidth]{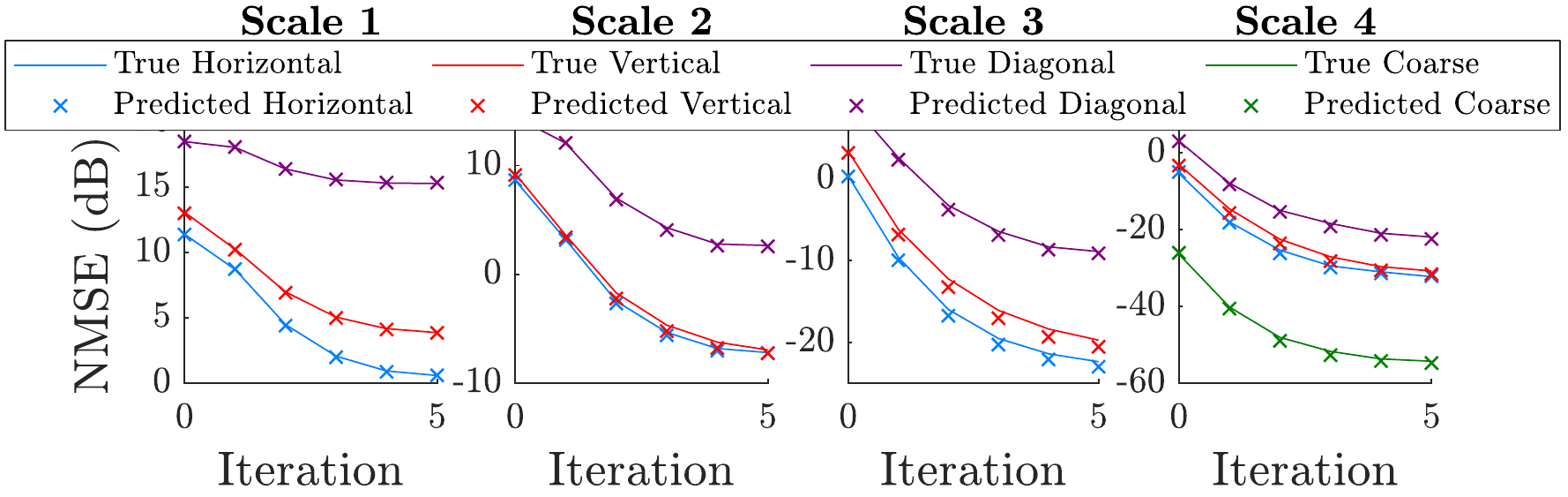}
		\label{fig:StateEvolution}
		\vspace{-16 pt}
	\end{subfigure}
	\caption{{\ninept{\bf State evolution.} (Top) An illustration of the effective error in the wavelet domain at iteration 2 of the algorithm: Each subband exhibits a different distribution and the elements within each subband appear uncorrelated. (Middle) Quantile-quantile plots of the empirical distribution within several of the effective error subbands: Linear plots indicate a Gaussian distribution. (Bottom) The empirical variances of the error within each subband alongside the predicted variances, $\tau_{k,s}$: The variances are accurately predicted.}}
	\label{fig:Finding}
\end{figure}

Within the iterations of VDAMP, the term $\bm{r}_{k}$ can be thought of as a noisy estimate of the true wavelet coefficients $\bm{w}_o=\bm{\Psi}\img_o$.
The difference between these two terms, $\boldsymbol{\nu}_k=\bm{r}_{k}-\bm{w}_o$, is known as the effective noise. 
Like VDAMP, D-VDAMP comes with a (weak) state evolution framework to predict the covariance of $\boldsymbol{\nu}_k$. The state evolution finding is stated below and is empirically supported in Figure~\ref{fig:Finding}.

\vspace{-6pt}
\begin{finding} The effective noise within each wavelet subband, $s$, follows a white Gaussian distribution with a variance equal to $\tau_{k,s}$ from line~\ref{algline:WaveletVariance}. 
\end{finding}
\vspace{-6pt}

\noindent Note that this ``state evolution'' differs from and is weaker than the typical AMP state evolution: The variances at iteration $k+1$ depends on the residual $\bm{z}_k$ from the previous iteration, not just the previous state evolution prediction. Accordingly, this state evolution is not self-contained and cannot be relied upon to analyze the noise sensitivity and convergence of the algorithm, as has been done with alternative AMP frameworks, e.g.,~\cite{donoho2011noise}. Instead, this state evolution only tells you the covariance of the effective noise at each iteration. 

\Subsection{Removing Colored Noise}

We propose a new CNN architecture which leverages the known covariance of the colored effective noise in order to remove it.  Our proposed architecture is 
a modified version of the DnCNN~\cite{zhang2017beyond} architecture where each intermediate feature map is concatenated with a $h\times w \times 13$ tensor where the $j^{th}$ slice of each tensor contains $h\times w$ copies of the standard deviation associated with the $j^{th}$ wavelet subband. (A Haar transform across 4 scales has 13 wavelet subbands.) The network architecture thereby provides all convolutional layers direct access to the standard deviations in each wavelet subband as extra channels.

Four copies of this CNN were trained to remove colored Gaussian noise that has a diagonal covariance matrix when represented in the wavelet domain. The networks were specialized for when the standard deviation of the noise across the 13 wavelet bands were uniformly distributed in the ranges 0--20, 20--50, 50--120, and 120--500, respectively. (Pixel values were in the range 0--255.)  
The training dataset was made of roughly 127\,000 different $48\times 48$ patches formed by cropping, scaling, flipping, and rotating 50 MRI images from the ``Stanford 2D FSE'' dataset available at \url{http://mridata.org/}, which is maintained by Miki Lustig and collaborators.  The networks was trained for 50 epochs using the ADAM optimizer with a learning rate of $0.001$ and batch size of 256. 


\Section{Simulation Results}\label{sec:Sims}

\begin{figure}[t]
	\centering
	\includegraphics[width=.08\textwidth]{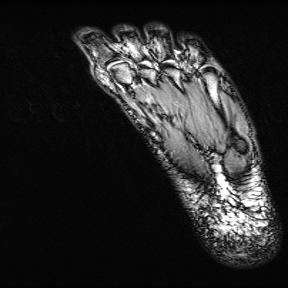}
	\includegraphics[width=.08\textwidth]{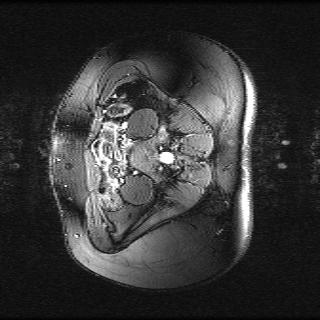}
	\includegraphics[width=.08\textwidth]{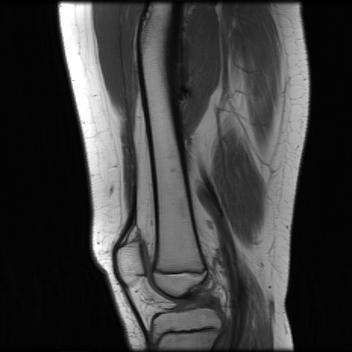}
	\includegraphics[width=.08\textwidth]{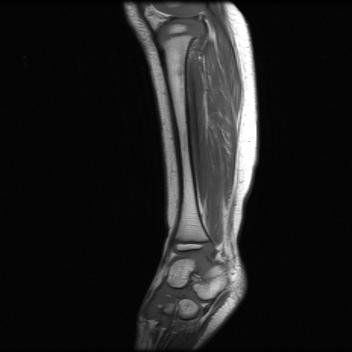}
	\includegraphics[width=.08\textwidth]{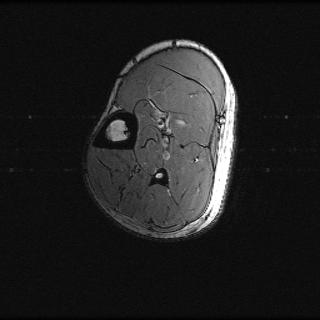}
	\vspace{5pt}
	\includegraphics[width=.08\textwidth]{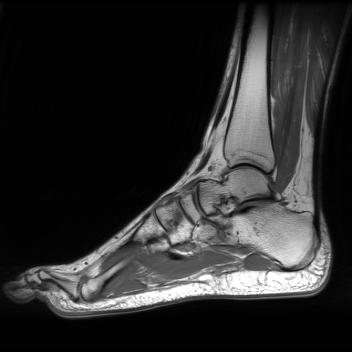}
	\includegraphics[width=.08\textwidth]{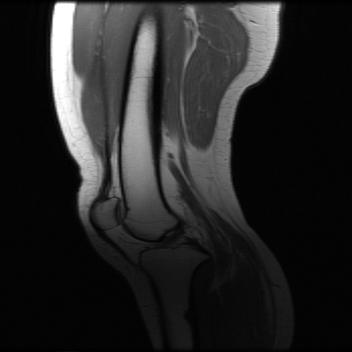}
	\includegraphics[width=.08\textwidth]{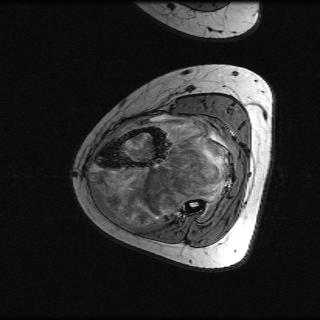}
	\includegraphics[width=.08\textwidth]{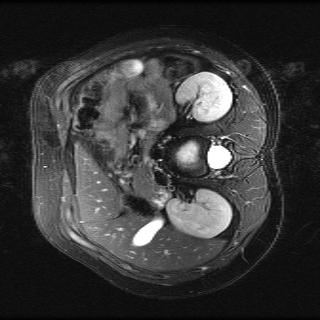}
	\includegraphics[width=.08\textwidth]{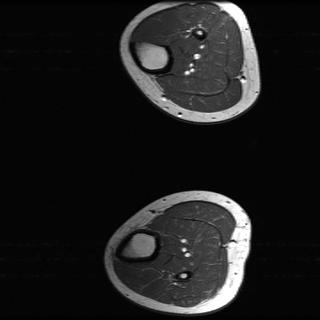}
	
	\vspace{-12 pt}
	\caption{{\ninept{\bf Testing dataset} from \url{http://mridata.org/}.}}
	\label{fig:TestImages}
\end{figure}

\begin{figure*}[t]
	\centering
	\begin{subfigure}[t]{.15\textwidth}
		\centering
		\begin{overpic}[width=\textwidth]{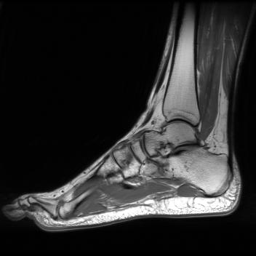}
			\put (0,88) {{\bf\raggedleft \textcolor{white}{Reference} }}
		\end{overpic}
		
		\begin{overpic}[width=\textwidth]{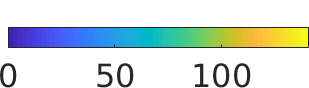}
		\end{overpic} 
	\end{subfigure}%
	~
	\begin{subfigure}[t]{.15\textwidth}
	\centering
	\begin{overpic}[width=\textwidth]{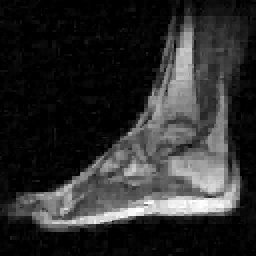}
		\put (0,88) {{\bf\raggedleft \textcolor{white}{VDAMP} }}
	\end{overpic} 

	\vspace{3pt}
	
	\begin{overpic}[width=\textwidth]{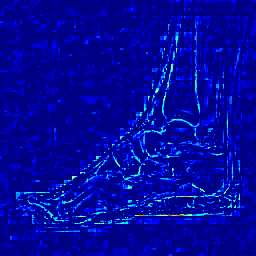}
		\put (0,90) {\textcolor{white}{PSNR = {\bf 23.5}}}
	\end{overpic} 
	\end{subfigure}%
	~
\begin{subfigure}[t]{.15\textwidth}
	\centering
	\begin{overpic}[width=\textwidth]{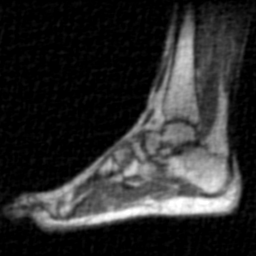}
		\put (0,88) {{\bf\raggedleft \textcolor{white}{TVAL3} }}
	\end{overpic} 
	
	\vspace{3pt}
	
	\begin{overpic}[width=\textwidth]{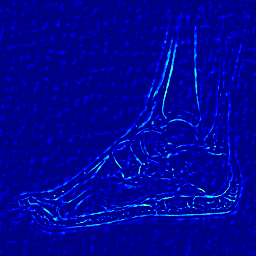}
		\put (0,90) {\textcolor{white}{PSNR = {\bf 26.9}}}
	\end{overpic} 
\end{subfigure}%
	~
\begin{subfigure}[t]{.15\textwidth}
	\centering
	\begin{overpic}[width=\textwidth]{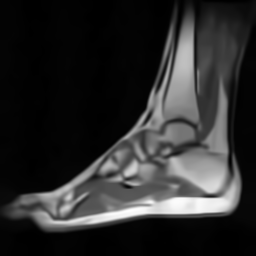}
		\put (0,80) {{\bf\raggedleft \textcolor{white}{\begin{tabular}{@{}l@{}} P\&P ADMM \\ (BM3D) \end{tabular}} }}
	\end{overpic} 
	
	\vspace{3pt}
	
	\begin{overpic}[width=\textwidth]{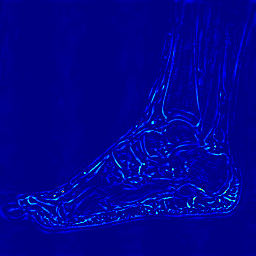}
		\put (0,90) {\textcolor{white}{PSNR = {\bf 29.0}}}
	\end{overpic} 
\end{subfigure}%
	~
\begin{subfigure}[t]{.15\textwidth}
	\centering
	\begin{overpic}[width=\textwidth]{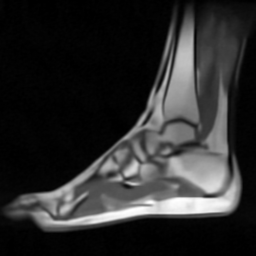}
		\put (0,80) {{\bf\raggedleft \textcolor{white}{\begin{tabular}{@{}l@{}} RED \\ (BM3D) \end{tabular}} }}
	\end{overpic} 

	\vspace{3pt}
		\begin{overpic}[width=\textwidth]{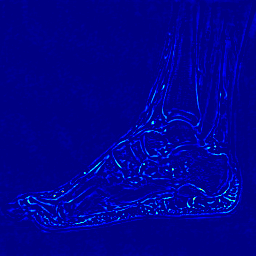}
		\put (0,90) {\textcolor{white}{PSNR = {\bf 29.4}}}
	\end{overpic} 
\end{subfigure}%
	~
\begin{subfigure}[t]{.15\textwidth}
	\centering
	\begin{overpic}[width=\textwidth]{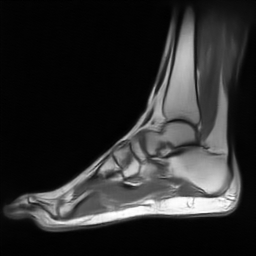}
		\put (0,80) {{\bf\raggedleft \textcolor{white}{\begin{tabular}{@{}l@{}} D-VDAMP \\ (Proposed) \end{tabular}} }}
	\end{overpic} 
	
	\vspace{3pt}
	
	\begin{overpic}[width=\textwidth]{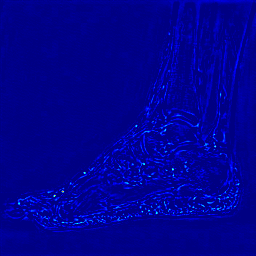}
		\put (0,90) {\textcolor{white}{PSNR = {\bf 31.0}}}
	\end{overpic} 
\end{subfigure}%
	\vspace{-8pt}
	\caption{{\ninept {\bf $\boldsymbol{256\times 256}$ reconstructions and error maps} for several algorithms from high SNR measurements with a sampling rate of $\frac{m}{n}=\frac{1}{16}$. The proposed D-VDAMP algorithm forms sharper and more accurate reconstructions than existing techniques.}}
	\label{fig:Reconstructions}
\end{figure*}

\begin{table*}[t]
	\ninept
	\vspace{-10pt}
	\parbox{\linewidth}{
	\centering
	\caption{Average PSNRs (dB) and runtimes (seconds) of $256 \times 256$ reconstructions of MRI images with high SNR (40 dB) variable density Fourier measurements with various sampling rates.}\vspace{-10pt}
	\begin{tabular}{lrrrrrrrr}
  \toprule
		\multirow{2}{*}{Method}&
		\multicolumn{2}{c}{$\frac{m}{n}=\frac{1}{16}$} & \multicolumn{2}{c}{$\frac{m}{n}=\frac{1}{12}$}& \multicolumn{2}{c}{$\frac{m}{n}=\frac{1}{8}$} & \multicolumn{2}{c}{$\frac{m}{n}=\frac{1}{4}$} \\
		\cmidrule{2-9} 
		& PSNR & Time& PSNR & Time & PSNR & Time & PSNR & Time  \\
		\midrule
		VDAMP &  25.2 &\textbf{0.2}& 26.1 &\textbf{0.2}& 27.6 &\textbf{0.3}& 32.3 &\textbf{0.5}\\
		TVAL3 &  27.7 &0.6& 28.3&0.6& 29.1 &0.7& 32.0&0.8 \\
		{P\&P ADMM (BM3D)} & 30.0 &112.0& 30.5 &114.7& 31.0 &114.3& 32.4 & 119.1 \\
		{P\&P ADMM (DnCNN, white)} & 27.6 &7.3& 28.9 &7.4& 30.7 &7.4& 33.6&7.5\\
		{RED (BM3D)} & 30.3 &284.1& 30.7 &282.9& 31.4 &281.8& 33.2&287.9 \\
		{RED (DnCNN, white)} & 28.8 &13.2& 29.9 &12.9& 31.2 &12.8& 34.3&12.4\\
		{D-VDAMP (BM3D)} & 28.4 &62.4& 30.2 &98.0& 32.5 &121.0& 39.9&157.0\\
		{D-VDAMP (DnCNN, white)} & 29.3 &2.9& 29.1 &4.1& 30.0 &4.4& 32.8&5.3\\
		{D-VDAMP (DnCNN, colored)} & 29.8 &2.9& 31.5 &4.1& 33.5 &4.4& 38.3 &5.3 \\
		{D-VDAMP (Proposed)} & \textbf{31.4} &3.1& \textbf{34.0} &4.1& \textbf{36.4} &5.9& \textbf{41.3} &6.1 \\
		\bottomrule
	\end{tabular}%
	\label{tab:highSNR}%
	}
\hfill
\parbox{\linewidth}{
	\centering
	\caption{Average PSNRs (dB) of $256 \times 256$ reconstructions of MRI images sampled with variable density Fourier measurements at a rate of $\frac{m}{n}=\frac{1}{8}$ under various SNRs.}\vspace{-10pt}
	\begin{tabular}{lccccccc}
		\toprule
		\multirow{2}{*}{Method}  & $\text{SNR}=25\text{dB}$ & $\text{SNR}=20\text{dB}$ & $\text{SNR}=15\text{dB}$ & $\text{SNR}=10\text{dB}$ & $\text{SNR}=5\text{dB}$\\
		\cmidrule{2-6}
		& PSNR &  PSNR & PSNR & PSNR& PSNR\\
		\midrule
		P\&P ADMM (BM3D) &   30.9 & 30.8 & 30.7 & \textbf{30.2} & \textbf{27.8} \\		
		RED (BM3D) & 31.4 & 31.1 & 30.6 & 29.3 & 25.0   \\
		D-VDAMP (BM3D) & 32.8 & 31.8 & 30.1 & 27.5 & 25.0 \\
		D-VDAMP (Proposed) &  \textbf{34.8} & \textbf{33.0} & \textbf{30.8} & 28.0 & 25.0\\
		\bottomrule
	\end{tabular}%
	\label{tab:lowSNR}%
	\vspace{-15pt}
}
\end{table*}%

\paragraph{Dataset} We test the algorithms on the $256\times 256$ MRI images presented in Figure~\ref{fig:TestImages}. (These images were not used to generate the training dataset.) For simplicity, our present simulations assume the signal is real-valued. The proposed method could be extended to deal with the more realistic complex-valued case by adding an extra channel to the input and output of the DnCNN denoiser, as is done in~\cite{ahmad2019plug}.

\paragraph{Measurements}
Our measurements are formed by adding i.i.d.~circular complex Gaussian noise to variable density sampled Fourier coefficients of the images~\cite{lustig2007sparse}. The variable density sampling strategy preferentially samples the low frequency components, which for MRI images generally contain more information. Extending the algorithm to other sampling strategies, such as Cartesian~\cite{ahmad2019plug}, remains future work.

\paragraph{Competition} We compare the proposed algorithm against the wavelet-sparsity based VDAMP algorithm~\cite{millard2020approximate} (with a 2-D Haar basis), the total variation based TVAL3 algorithm~\cite{TVAL3}, and two plug-and-play algorithms: P\&P ADMM~\cite{PnPADMM} and RED~\cite{RED}. The latter two algorithms represent the state-of-the-art on this task~\cite{ahmad2019plug}. Note that, while they can offer excellent performance when trained on a fixed measurement matrix, unrolled methods like ADMM-Net~\cite{sun2016deep} do not generalize across measurement patterns, and so are not compared against here.

We test both plug and play algorithms with both the classical BM3D denoising algorithm~\cite{BM3D} and the CNN-based DnCNN denoising algorithm~\cite{zhang2017beyond}, trained with white Gaussian noise. 
Similarly, we test the D-VDAMP algorithm with BM3D, a DnCNN network trained with white noise, a DnCNN network trained with colored noise, and our proposed, modified DnCNN network trained with colored noise. 

\paragraph{Algorithm Parameters}
VDAMP is run for 30 iterations, TVAL3 is run for 300 iterations. P\&P ADMM and RED are both run for 200 iterations. D-VDAMP is run for 10 iterations or until the predicted power of the effective noise, as given by the sum of the elements of $\boldsymbol{\tau}_k$, increases.

Both P\&P ADMM and RED use denoisers whose parameters were optimized to denoise noise with a standard deviation between 20--50 with DnCNN or 25 with BM3D (among the four networks, this one worked the best). 
The RED loss is minimized using proximal gradient descent with adaptive step sizes, as implemented by FASTA~\cite{FASTA}, and with a proximal mapping step that applies the denoiser only once. P\&P ADMM's $\rho$ and $\gamma$ parameters, as defined in~\cite{chan2016plug}, are both set to 1. RED's $\lambda$ parameter is also set to 1. D-VDAMP's $\gamma$ parameter is set to $0.75$. All the P\&P algorithms denoise only the real part of the signal. P\&P ADMM and RED simply zero out the imaginary part while D-VDAMP instead scales the imaginary part by $0.1$, so as to maintain a nonzero average partial derivative.

\paragraph{Results}
Table~\ref{tab:highSNR} demonstrates that when dealing with high SNR measurements, the proposed algorithm substantially outperforms the state-of-the-art. It outperforms wavelet based methods by nearly 10 dB and uniformly outperforms other P\&P methods by at least 1 dB. As illustrated in Figure~\ref{fig:Reconstructions}, the proposed algorithm results in reconstructions with fewer artifacts and sharper edges.

However, as noted in~\cite{millard2020approximate}, the density compensated gradient step of VDAMP (line~\ref{algline:pseudodata}), which multiplies $\bm{z}_k$ by $\bm{P}^{-1}$, makes D-VDAMP sensitive to measurement noise. Accordingly, as demonstrate in Table~\ref{tab:lowSNR}, at lower SNRs existing P\&P algorithms outperform D-VDAMP.

\Section{Discussion}

A chief criticism against many learning-based reconstruction methods is that they can produce realistic looking images, even when the measurements contain little information about the actual scene~\cite{gottschling2020troublesome}. These images can lead practitioners to trust the reconstruction, even when they should not. 
Fortunately, because the last step of D-VDAMP is equivalent to removing colored Gaussian noise with known covariance from the ground-truth signal, Stein's unbiased risk estimate can be combined with D-VDAMP to produce heat maps of the expected squared error per pixel associated with the reconstruction~\cite{deledalle2012non,edupuganti2020uncertainty,KaosPaper}. This allows practitioners to judge which portions of the reconstruction can be trusted.

\Section{References}
\renewcommand{\section}[1]{}

{
\ninept
\bibliographystyle{IEEEbib}
\bibliography{myref}
}
\end{document}